\documentclass[preprint,prd,showpacs,superscriptaddress,preprintnumbers,nofootinbib,fleqn,floatfix]{revtex4}

\usepackage{amsmath}
\usepackage{amssymb}
\usepackage{graphicx}
\usepackage{dcolumn}
\usepackage{enumerate}
\usepackage{axodraw}

\begin{document}


\preprint{RIKEN-TH 182}


\title{ 
 Tenth-order lepton $g\!-\!2$:  \\
Contribution from diagrams containing
a sixth-order light-by-light-scattering  subdiagram
internally
}

\author{Tatsumi Aoyama} 
\affiliation{Faculty of Science, Nagoya University, Nagoya, 
Aichi 464-8602, Japan} 

\author{Katsuyuki Asano}
\affiliation{Department of Physics, Nagoya University, Nagoya, 
Aichi 464-8602, Japan}

\author{Masashi Hayakawa}
\affiliation{Department of Physics, Nagoya University, Nagoya, 
Aichi 464-8602, Japan}  

\author{Toichiro Kinoshita} 
\affiliation{Laboratory for Elementary-Particle Physics, 
Cornell University, Ithaca, New York 14853, USA}  

\author{Makiko Nio} 
\affiliation{Theoretical Physics Laboratory, Nishina Center, 
RIKEN, Wako, 351-0198, Japan} 

\author{Noriaki Watanabe} 
\affiliation{Department of Physics, Nagoya University, Nagoya, 
Aichi 464-8602, Japan}
 
\date{\today}

\begin{abstract} 
 This paper reports the result of our evaluation of the tenth-order QED
correction to the lepton $g\!-\!2$ from Feynman diagrams which have 
sixth-order light-by-light-scattering subdiagrams,
none of whose vertices couple to the external magnetic field.
 The gauge-invariant set of these diagrams, called Set II(e),
consists of 180 vertex diagrams.
 In the case of the electron $g\!-\!2$ ($a_e$),
where the light-by-light subdiagram consists of the electron loop,
the contribution to $a_e$ is found to be
$- 1.344\ 9\ (10) \left(\alpha /\pi\right)^5$. 
 The contribution of the muon loop to $a_e$ is 
$- 0.000~465\ (4) \left(\alpha /\pi\right)^5$. 
 The contribution of the tau-lepton loop is about two orders of
magnitudes smaller than that of the muon loop and hence negligible.
 The sum of all of these contributions to $a_e$ is
$- 1.345\ (1) \left(\alpha /\pi\right)^5$. 
 We have also evaluated the contribution of Set II(e) to
the muon $g\!-\!2$ ($a_\mu$).
 The contribution to $a_\mu$ from the electron loop is 
$3.265\ (12) \left(\alpha /\pi\right)^5$,
while the contribution of the tau-lepton loop is
$-0.038~06\ (13) \left(\alpha /\pi\right)^5$.
 The total contribution to $a_\mu$,
which is the sum of these two contributions
and the mass-independent part of $a_e$,
is $1.882\ (13) \left(\alpha /\pi\right)^5$.
\end{abstract} 

\pacs{13.40.Em,14.60.Cd,14.60Ef,12.20.Ds} 

\maketitle 

\section{Introduction}
\label{sec:intro}

 The anomalous magnetic moment $g-2$ of the electron 
has provided the most stringent test of  
the validity of quantum electrodynamics, QED.
The experimental value with the least uncertainty 
is the one obtained by the Harvard group in 2008 ($a \equiv \frac{g-2}{2}$)
\cite{Hanneke:2008tm}
\begin{eqnarray}
 a_e({\rm HV08}) &=&
 1\ 159\ 652\ 180.73\ (28) \times 10^{-12}\,.
  \label{eq:Harvard_exp08}
\end{eqnarray}
 To confront the prediction of the standard model 
with this measurement
the hadronic contribution up to the order $\alpha^3$ 
\cite{Hagiwara:2006jt,Jegerlehner:2009ry,Davier:2009zi,Krause:1996rf,
Melnikov:2003xd,Bijnens:2007pz,Prades:2009tw,Nyffeler:2009tw}, 
the electroweak contribution up to the two-loop order
\cite{Czarnecki:1995sz,Knecht:2002hr,Czarnecki:2002nt}, 
and the QED radiative correction up to the eighth order
must be taken into account 
\cite{Kinoshita:2005zr,Aoyama:2007dv,Aoyama:2007mn}.
 In order to match or exceed further 
improvement in the accuracy 
of the experimental value, it is necessary to evaluate 
the tenth-order QED radiative correction to $g - 2$.
 To meet this challenge we launched several years ago the project to compute  
all 12672 Feynman diagrams that contribute to the tenth-order $a_e$ 
\cite{Aoyama:2005kf,Kinoshita:2005sm}.

 The most difficult to evaluate is the gauge-invariant set, called Set V,
which consists of 6354 diagrams that have no virtual lepton loop.
 To deal with this set systematically
we have developed an automatic code-generating algorithm {\sc gencode}{\it N}
\cite{Aoyama:2005kf,Aoyama:2007bs}.
 We now have
{\sc fortran} codes for all diagrams of Set V generated by {\sc gencode}{\it N}.
Numerical evaluation of these integrals is in progress at present.

\begin{figure}
        \includegraphics[scale=1.0]{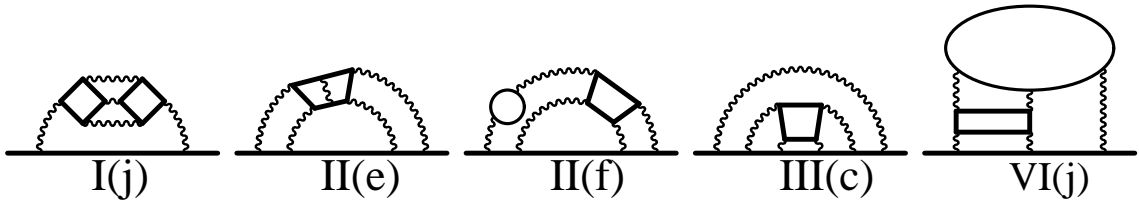}
\caption{Representative diagrams of gauge-invariant sets which contain
an internal light-by-light-scattering subdiagram. 
\label{fig:Xllin}
}
\end{figure}

 Meanwhile, we have also made steady progress in the evaluation 
of other types of tenth-order diagrams, 
and have published some of the results 
\cite{Kinoshita:2005sm,Aoyama:2008gy,Aoyama:2008hz}.
 At the tenth order there appear five gauge-invariant sets 
of diagrams, called Set I(j), Set II(e), Set II(f), Set III(c),
and Set VI(j),
which contain light-by-light-scattering subdiagram(s) internally,
i.e., none of whose vertices is the external vertex  
\cite{Aoyama:2005kf,Kinoshita:2005sm}.
(See Fig.~\ref{fig:Xllin}.)

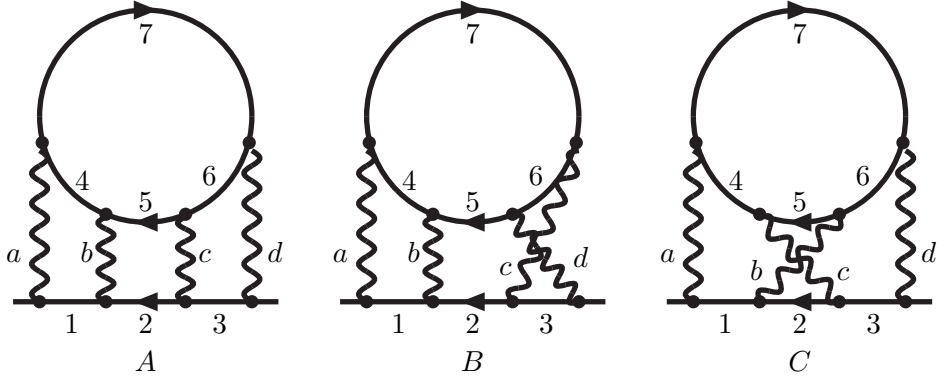
\begin{figure}[t]
\begin{picture}(120,160)(-60,-80) 
\SetWidth{2.0} 
\ArrowArcn(0,20)(40,0,-180)  
\ArrowArcn(0,20)(40,-180,-360) 
\ArrowLine(50,-50)(-50,-50) 
\Photon(-40,-50)(-40,7){3}{5}
\Photon(40,-50)(40,7){3}{5}
\Photon(-15,-50)(-15,-18){3}{3.5}
\Photon(15,-50)(15,-18){3}{3.5} 
\Vertex(-40,-50){2.5} 
\Vertex(40,-50){2.5} 
\Vertex(-15,-50){2.5} 
\Vertex(15,-50){2.5} 
\Text(-28,-55)[t]{$1$}  
\Text(0,-55)[t]{$2$}  
\Text(28,-55)[t]{$3$}  
\Vertex(-39,10){2.5} 
\Vertex(39,10){2.5} 
\Vertex(-15,-17){2.5} 
\Vertex(15,-17){2.5} 
\SetWidth{1.5} 
\Text(-47,-35)[br]{$a$}
\Text(-20,-35)[br]{$b$}
\Text(20,-35)[bl]{$c$}
\Text(46,-35)[bl]{$d$}
\Text(-24,-1)[t]{$4$} 
\Text(0,-9)[t]{$5$} 
\Text(24,1)[t]{$6$} 
\Text(0,55)[t]{$7$}
\Text(0,-69)[t]{$A$}
\end{picture} 
\begin{picture}(120,160)(-60,-80) 
\SetWidth{2.0} 
\ArrowArcn(0,20)(40,0,-180)  
\ArrowArcn(0,20)(40,-180,-360) 
\ArrowLine(50,-50)(-50,-50) 
\Photon(-40,-50)(-40,7){3}{5}
\Photon(-15,-50)(-15,-18){3}{3.5}
\Photon(40,-50)(15,-18){3}{5}
\Photon(15,-50)(40,7){3}{5.5} 
\Vertex(-40,-50){2.5} 
\Vertex(40,-50){2.5} 
\Vertex(-15,-50){2.5} 
\Vertex(15,-50){2.5} 
\Text(-28,-55)[t]{$1$}
\Text(0,-55)[t]{$2$}  
\Text(28,-55)[t]{$3$}  
\Vertex(-39,10){2.5} 
\Vertex(39,10){2.5} 
\Vertex(-15,-17){2.5} 
\Vertex(15,-17){2.5} 
\SetWidth{1.5} 
\Text(-47,-35)[br]{$a$}
\Text(-20,-35)[br]{$b$}
\Text(10,-40)[bl]{$c$}
\Text(38,-37)[bl]{$d$}
\Text(-24,-1)[t]{$4$} 
\Text(0,-9)[t]{$5$} 
\Text(24,1)[t]{$6$} 
\Text(0,55)[t]{$7$}
\Text(0,-69)[t]{$B$}
\end{picture} 
\begin{picture}(120,160)(-60,-80) 
\SetWidth{2.0} 
\ArrowArcn(0,20)(40,0,-180)  
\ArrowArcn(0,20)(40,-180,-360) 
\ArrowLine(50,-50)(-50,-50) 
\Photon(-40,-50)(-40,7){3}{5}
\Photon(40,-50)(40,7){3}{5}
\Photon(-15,-50)(15,-18){3}{4}
\Photon(15,-50)(-15,-18){3}{4} 
\Vertex(-40,-50){2.5} 
\Vertex(40,-50){2.5} 
\Vertex(-15,-50){2.5} 
\Vertex(15,-50){2.5} 
\Text(-28,-55)[t]{$1$}  
\Text(0,-55)[t]{$2$}  
\Text(28,-55)[t]{$3$}  
\Vertex(-39,10){2.5} 
\Vertex(39,10){2.5} 
\Vertex(-15,-17){2.5} 
\Vertex(15,-17){2.5} 
\SetWidth{1.5} 
\Text(-47,-35)[br]{$a$}
\Text(-14,-42)[br]{$b$}
\Text(14,-42)[bl]{$c$}
\Text(46,-35)[bl]{$d$}
\Text(-24,-1)[t]{$4$} 
\Text(0,-9)[t]{$5$} 
\Text(24,1)[t]{$6$} 
\Text(0,55)[t]{$7$}
\Text(0,-69)[t]{$C$}
\end{picture} 
\begin{center}
\caption{
 Eighth-order self-energy Feynman diagrams $LL8$
that contain the fourth-order light-by-light-scattering subdiagram
internally.
 In Ref.~\cite{Kinoshita:1981ww}, 
$A$, $B$ and $C$ are called ``LLJ'', ``LLL'' and ``LLK'', respectively.
 The vertex diagrams are obtained by inserting a single QED vertex
into one of the lepton lines $1,\,2$ or $3$ 
of $G$ ($G = A,\,B,\,C$).
\label{fig:8thCILbyL}
}
\end{center}
\end{figure}

 Feynman diagrams
containing a light-by-light-scattering subdiagram internally
appear for the first time in the eighth-order QED correction
to the lepton $g\!-\!2$.
 Figure~\ref{fig:8thCILbyL} shows the eighth-order self-energy diagrams
with the fourth-order internal light-by-light-scattering subdiagrams.
 Vertex diagrams relevant to lepton $g\!-\!2$ can be obtained by 
inserting a single external QED vertex 
into one of open lepton lines labeled $1,\,2,\,3$
of individual diagrams of Figure~\ref{fig:8thCILbyL}.

 The diagrams of Set I(j) are those involving two 
fourth-order light-by-light-scattering subdiagrams, both internal,
which have been evaluated  and published recently \cite{Aoyama:2008gy}.
 The diagrams of other subsets 
are obtained by adding $O(\alpha)$ correction to 
those of Fig.~\ref{fig:8thCILbyL}.
The Set II(f) consists of diagrams obtained by inserting 
a second-order vacuum-polarization loop into one of internal photon lines
of the diagrams of Fig.~\ref{fig:8thCILbyL} in all possible ways.
 They have been evaluated 
by a simple modification of the {\sc fortran} codes 
developed previously for the eighth-order work.
The result was published in Ref.~\cite{Kinoshita:2005sm}. 
 The diagrams of Set III(c)
are obtained by attaching a virtual photon line 
to the open lepton path
of the individual diagrams of Fig.~\ref{fig:8thCILbyL} 
in all possible ways.
Evaluation of this set is in progress.
 The diagrams of Set VI(j)
contain two light-by-light-scattering subdiagrams, one of which is internal,
while the other is external.
The numerical result of the Set VI(j) was published
in Ref.~\cite{Kinoshita:2005sm}.

 The diagrams of Set II(e)
are obtained by attaching both ends of a virtual photon line 
to the lepton loop 
of the individual diagrams of Fig.~\ref{fig:8thCILbyL} 
in all possible ways, forming
the sixth-order internal light-by-light-scattering subdiagram.
 This paper reports the result of our work on 
Set II(e), which consists of $180$ Feynman vertex diagrams.

 The paper is organized as follows.
 Section \ref{sec:comp_proc} describes the strategy we have adopted 
for the numerical study.
 The renormalization is set up 
so that ultraviolet divergences can be subtracted away
without introducing spurious infrared divergence.
 Section \ref{sec:int_renorm} gives the results of our numerical work
which covers the contributions of all diagrams of Set II(e) to 
$a_e$ and $a_\mu$.
 Section \ref{sec:conclusion} is devoted to the discussion and summary.

\section{Computational procedure}
\label{sec:comp_proc}
 This section describes the strategy for computing the diagrams of Set II(e).
 We denote the contribution of Set II(e) to the magnetic moment
of the lepton $l$
induced by the virtual loop of lepton $l^\prime$ as 
\begin{eqnarray}
 a_l\left({\rm II(e)},\,l^\prime\right) 
 =
 A_l\left({\rm II(e)},\,l^\prime\right)
 \left(\frac{\alpha}{\pi}\right)^5\,,
\end{eqnarray}
where the lower case "{\it a}" includes the factor 
$\left(\frac{\alpha}{\pi}\right)^5$ while the upper case "{\it A}" does not.
 Recall that $a_l\left({\rm II(e)},\,l^\prime = l\right)$ 
and $A_l\left({\rm II(e)},\,l^\prime = l\right)$
are independent of $l$ and called mass-independent contributions.
 $A_l\left({\rm II(e)},\,l^\prime \ne l\right)$
depends only on the mass ratio $m_{l^\prime} /m_l$.
 We use the values found in Ref.~\cite{Amsler:2008zzb} 
for lepton masses.

\begin{figure}[b]
\begin{center} 
\begin{picture}(120,160)(-60,-80) 
\SetWidth{2.0} 
\ArrowArcn(0,20)(40,0,-180)  
\ArrowArcn(0,20)(40,-180,-360) 
\ArrowLine(50,-50)(-50,-50) 
\Photon(-40,-50)(-40,7){3}{5}
\Photon(40,-50)(40,7){3}{5}
\Photon(-15,-50)(-15,-18){3}{3.5}
\Photon(15,-50)(15,-18){3}{3.5} 
\Vertex(-40,-50){2.5} 
\Vertex(40,-50){2.5} 
\Vertex(-15,-50){2.5} 
\Vertex(15,-50){2.5} 
\Text(-28,-55)[t]{$1$}  
\Text(0,-55)[t]{$2$}  
\Text(28,-55)[t]{$3$}  
\Vertex(-39,10){2.5} 
\Vertex(39,10){2.5} 
\Vertex(-15,-17){2.5} 
\Vertex(15,-17){2.5} 
\SetWidth{1.5} 
\Text(-48,-35)[br]{$a$}
\Text(-20,-35)[br]{$b$}
\Text(21,-35)[bl]{$c$}
\Text(47,-35)[bl]{$d$}
\SetWidth{2}
\PhotonArc(-40,10)(20,-55,88){2}{5} 
\Vertex(-39,28){2.5}
\Vertex(-27.5,-9){2.5}
\SetWidth{1.5}
\Text(-14,10)[bl]{$e$} 
\Text(-34,23)[t]{$8$} 
\Text(-31,7)[t]{$9$} 
\Text(-20,-5)[t]{$4$} 
\Text(0,-9)[t]{$5$} 
\Text(24,1)[t]{$6$} 
\Text(0,55)[t]{$7$} 
\Text(0,-68)[t]{$A47$}
\end{picture} 
\begin{picture}(120,160)(-60,-80) 
\SetWidth{2.0} 
\ArrowArcn(0,20)(40,0,-180)  
\ArrowArcn(0,20)(40,-180,-360) 
\ArrowLine(50,-50)(-50,-50) 
\Photon(-40,-50)(-40,7){3}{5}
\Photon(40,-50)(40,7){3}{5}
\Photon(-15,-50)(-15,-18){3}{3.5}
\Photon(15,-50)(15,-18){3}{3.5} 
\Vertex(-40,-50){2.5} 
\Vertex(40,-50){2.5} 
\Vertex(-15,-50){2.5} 
\Vertex(15,-50){2.5} 
\Text(-28,-55)[t]{$1$}  
\Text(0,-55)[t]{$2$}  
\Text(28,-55)[t]{$3$}  
\Vertex(-39,10){2.5} 
\Vertex(39,10){2.5} 
\Vertex(-15,-17){2.5} 
\Vertex(15,-17){2.5} 
\SetWidth{1.5} 
\Text(-46,-35)[br]{$a$}
\Text(-20,-35)[br]{$b$}
\Text(21,-35)[bl]{$c$}
\Text(47,-35)[bl]{$d$}
\SetWidth{2}
\PhotonArc(-29,-7)(6,-55,147){2}{3.5} 
\Vertex(-33,-3){2.5}
\Vertex(-23,-12.5){2.5}
\SetWidth{1.5}
\Text(-21,-1.6)[bl]{$e$} 
\Text(-33,10)[t]{$8$} 
\Text(-31,-9)[t]{$9$} 
\Text(-22,-16)[t]{$4$} 
\Text(3,-9)[t]{$5$} 
\Text(24,1)[t]{$6$} 
\Text(0,55)[t]{$7$} 
\Text(0,-68)[t]{$A44$}
\end{picture} 
\begin{picture}(120,160)(-60,-80) 
\SetWidth{2.0} 
\ArrowArcn(0,20)(40,0,-180)  
\ArrowArcn(0,20)(40,-180,-360) 
\ArrowLine(50,-50)(-50,-50) 
\Photon(-40,-50)(-40,7){3}{5}
\Photon(40,-50)(40,7){3}{5}
\Photon(-15,-50)(-15,-18){3}{3.5}
\Photon(15,-50)(15,-18){3}{3.5} 
\Vertex(-40,-50){2.5} 
\Vertex(40,-50){2.5} 
\Vertex(-15,-50){2.5} 
\Vertex(15,-50){2.5} 
\Text(-28,-55)[t]{$1$}  
\Text(0,-55)[t]{$2$}  
\Text(28,-55)[t]{$3$}  
\Vertex(-39,10){2.5} 
\Vertex(39,10){2.5} 
\Vertex(-15,-17){2.5} 
\Vertex(15,-17){2.5} 
\SetWidth{1.5} 
\Text(-46,-35)[br]{$a$}
\Text(-20,-35)[br]{$b$}
\Text(21,-35)[bl]{$c$}
\Text(47,-35)[bl]{$d$}
\SetWidth{2}
\Photon(0,-20)(0,58){4}{7} 
\Vertex(0,-20){2.5}
\Vertex(0,58){2.5}
\SetWidth{1.5}
\Text(-12,10)[bl]{$e$} 
\Text(-25,2)[t]{$4$} 
\Text(-7,-22)[t]{$5$} 
\Text(7,-22)[t]{$8$} 
\Text(24,1)[t]{$6$} 
\Text(28,35)[t]{$7$} 
\Text(-28,35)[t]{$9$} 
\Text(0,-68)[t]{$A57$}
\end{picture} 
\caption{Representative diagrams of Set II(e).
$A47$ and $A44$ involve a second-order vertex subdiagram and 
a second-order self-energy subdiagram, respectively.
\label{fig:setII(e)}
}
\end{center}
\end{figure}
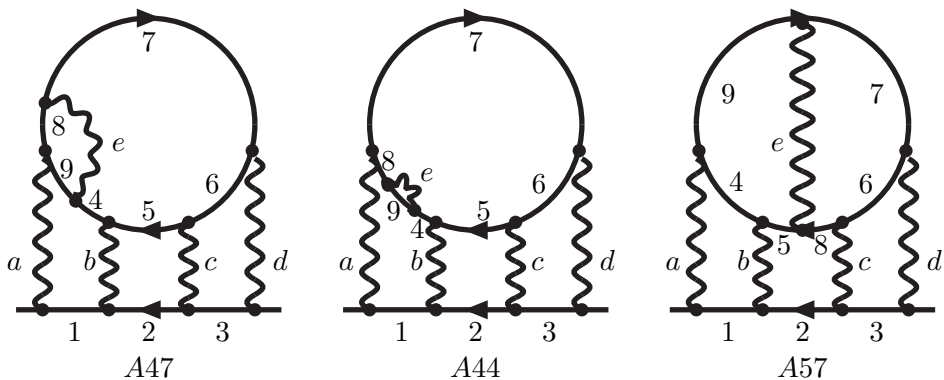

 As explained in Sec.~\ref{sec:intro},
the diagrams of Set II(e) are obtained by attaching
an internal photon line to the lepton loop of the individual diagrams 
of Fig.~\ref{fig:8thCILbyL} in all possible ways.
Let us denote the diagrams of Set II(e) as $Gij$ 
by specifying (i) the base eighth-order diagram $G$, where $G$ is
one of $A$, $B$, or $C$ of Fig.~\ref{fig:8thCILbyL}, 
and (ii) a pair $(i,\,j)$ of lepton lines of the closed loop 
($4 \le i \le j \le 7$) to which 
an additional internal photon line is attached.
 For instance, 
the insertion of two QED vertices into 
the middle of the lines $4$ and $7$ of the diagram $A$
of Fig.~\ref{fig:8thCILbyL} 
and the introduction of a virtual photon line 
which connects these vertices 
produces the tenth-order diagram called $A47$.
 Representative diagrams of Set II(e) are shown in Figure~\ref{fig:setII(e)}.
 The charge conjugation and time reversal symmetries of QED 
help us to reduce the number of independent amplitudes.
 For instance, $A67$ gives the same contribution to $g\!-\!2$ as $A47$.
 Recall also that the diagram in which the lepton loop runs in the
opposite direction
gives the same contribution as the original one.
 In this way, we obtain a complete set of independent diagrams:
\begin{eqnarray}
 &&
 \displaystyle{
  A44\,[4],\,A55\,[2],\,A77\,[2],\,A47\,[4],\,A45\,[4],\,
  A46\,[2],\,A57\,[2],\,
 }\nonumber\\
 &&
 \displaystyle{
  B44\,[4],\,B55\,[4],\,B47\,[4],\,B45\,[4],\,
  B46\,[2],\,B57\,[2],\,
 }\nonumber\\
 &&
 \displaystyle{
  C44\,[4],\,C55\,[2],\,C77\,[2],\,C47\,[4],\,C45\,[4],\,
  C46\,[2],\,C57\,[2],\,
 }\label{eq:indpndntDiag_setIIe}
\end{eqnarray}
where the number in the brackets accounts for the symmetry factor 
for each diagram as well as the doubling due to two directions
that a lepton loop takes.

 Thus far no one has succeeded
in evaluating the diagrams of Set II(e) analytically. 
 We resort to the numerical means utilizing
the parametric integral formulation 
\cite{Cvitanovic:1974uf,Cvitanovic:1974sv,Aoyama:2005kf,Aoyama:2007bs}.
 The evaluation of $g\!-\!2$ can be simplified significantly
by focusing on the quantity associated with the self-energy diagram $Gij$, 
such as the magnetic moment amplitude $M_{Gij}$, 
using the Ward-Takahashi identity which relates
the regularized self-energy function  $\Sigma_{Gij}(p)$ 
of the diagram $Gij$ to 
the sum $\Lambda_{Gij}(p,\,q)$ of the contributions
from the regularized vertex diagrams obtained by inserting 
a QED vertex into $Gij$ in all possible ways \cite{Cvitanovic:1974um}.

The next step is to renormalize the integrals on the computer,
which we carry out by subtractive renormalization.
 Since the bare amplitudes of individual diagrams 
have different structures of UV singularities,
the numerical subtraction of UV singularities 
must be carried out for each diagram separately.
Our aim is to construct subtraction terms that 
(i) share the same UV singularity 
as the integrand of the bare amplitude in the common Feynman parameter
space,
and
(ii) do not introduce spurious IR singularities.

 The second point is not a trivial requirement. 
 For instance,
the usual on-shell second-order vertex renormalization constant
contains an IR divergence.  In general the subtraction term
constructed under the usual on-shell renormalization condition
introduces an IR singularity that is not present 
in the bare amplitude.
To avoid this problem we perform the renormalization in two steps.
 The first step is 
to construct the UV-finite amplitude $\Delta M_{Gij}$ 
in which only the UV-divergent part of 
the corresponding on-shell vertex (or self-energy) term is subtracted, 
leaving out the UV-finite piece unsubtracted.
 We call this step an {\it intermediate renormalization}.
 The second step is to carry out the finite {\it residual} renormalization
to account for the difference between the intermediate renormalization
and the usual on-shell renormalization.
 The IR-divergent parts of the usual on-shell renormalization constants
appear in the second step but
cancel out when summed over the entire gauge-invariant set.

 The subtraction terms of $\Delta M_{Gij}$ are constructed as follows.
 The UV singularities associated 
with the second-order vertex and self-energy subdiagrams are subtracted 
via ${\sf K}$-operation, retaining the Feynman cut-off 
until UV divergences cancel out by renormalization \cite{Cvitanovic:1974sv}.
 The UV singularities of the  
light-by-light-scattering ({\it l-l})  loops are subtracted 
while maintaining the Pauli-Villars regularization in order to
avoid dealing with divergent hence undefined quantities.
 The Pauli-Villars mass is sent to infinity only after 
renormalization is carried out.
 Note that not only the usual on-shell renormalization 
but also the intermediate renormalization 
are defined on the mass-shell insofar as it is IR-safe.
 To avoid confusion let us call the usual on-shell renormalization as 
the {\it full renormalization} henceforth.

 Let us illustrate our renormalization procedure 
taking $A47$ of Fig.~\ref{fig:setII(e)} as an example.
 $A47$ has four UV-divergent subdiagrams
which can be identified by the sets of lepton lines involved:
\begin{eqnarray}
 &
 \displaystyle{
  S_1 = \left\{ 1,\,2,\,4,\,5,\,6,\,7,\,8,\,9 \right\},\,\quad
  S_2 = \left\{ 2,\,3,\,4,\,5,\,6,\,7,\,8,\,9 \right\},\,
 }&\nonumber\\
 &
 \displaystyle{
  S_3 = \left\{ 4,\,5,\,6,\,7,\,8,\,9 \right\},\,\quad
  S_4 = \left\{ 8,\,9 \right\}\,.
 }&
\end{eqnarray}
 Both subdiagrams $S_1$ and $S_2$ are the eighth-order vertex subdiagrams,
$S_3$ is the sixth-order {\it l-l} subdiagram, and $S_4$ is the
second-order vertex subdiagram.
 Each UV subtraction term of $\Delta M_{Gij}$ is associated with 
a Zimmermann's forest. 
 $A47$ has 11 normal forests. 
Let ${\sf C}_S$ denote 
the operator which extracts the full renormalization constant 
of the subset $S$ from $M_{Gij}$, and let
${\sf K}_S$ denote 
the operator which  extracts the UV singularity of the subset $S$ by
the intermediate renormalization defined by the ${\sf K}$ operation, 
respectively.  Then the UV-finite quantity $\Delta M_{A47}$ is defined by
\begin{eqnarray}
 \Delta M_{A47} &=&
 M_{A47} 
 - {\sf C}_{S_1} M_{A47} - {\sf C}_{S_2} M_{A47} 
 - {\sf C}_{S_3} M_{A47} - {\sf K}_{S_4} M_{A47}\nonumber\\
 &&
 + {\sf K}_{S_4} {\sf C}_{S_3} M_{A47}
 + {\sf K}_{S_4} {\sf C}_{S_1} M_{A47}
 + {\sf K}_{S_4} {\sf C}_{S_2} M_{A47}\nonumber\\
 &&
 + {\sf C}_{S_3} {\sf C}_{S_1} M_{A47}
 + {\sf C}_{S_3} {\sf C}_{S_2} M_{A47}\nonumber\\
 &&
 - {\sf K}_{S_4} {\sf C}_{S_3} {\sf C}_{S_1} M_{A47}
 - {\sf K}_{S_4} {\sf C}_{S_3} {\sf C}_{S_2} M_{A47}\,.
\label{eq:subtraction}
\end{eqnarray}
 Expression of $\Delta M_{Gij}$ 
for other $Gij$ in Eq.~(\ref{eq:indpndntDiag_setIIe})
can be written down similarly.

 It is by definition that all subtraction terms 
on the right-hand side of Eq.~(\ref{eq:subtraction}) are factorizable.
 For instance, the operator ${\sf C}_{S_1}$ acting on $M_{\rm A47}$ 
produces the product of $L_{S_1}$ and $M_2$:
\begin{eqnarray}
 {\sf C}_{S_1} M_{\rm A47} &=& L_{S_1}\,M_2\,,
\end{eqnarray}
where $L_{S_1}$ is 
the full eighth-order vertex renormalization constant of the diagram 
that contains the sixth-order light-by-light-scattering subdiagram, 
and $M_2 = a_2 = \frac{1}{2}$ is the second-order lepton $g\!-\!2$.
 Of course this equation is meaningless unless it is regularized.
 $L_{S_1}$ and $M_2$ can be expressed as regularized integrals 
in the parametric integral formulation \cite{Cvitanovic:1974uf}
on two separate Feynman parameter spaces 
with constraints $\sum_{i\,:\,{\rm all\ lines} \in S_1} x_i = 1$, 
$\sum_{k = 3,\,d} y_k = 1$, where $y_d$ is the Feynman parameter 
associated with the photon $d$.

 A  manipulation similar to 
that of Sec.~III D of Ref.~\cite{Aoyama:2005kf} 
expresses $L_{S_1}\,M_2$ as an integral 
over the single Feynman parameter space 
with $\sum_{j \in A47} z_j = 1$.
 With this form of integrand of $L_{S_1}\,M_2$ the pointwise
subtraction of the overall UV divergence of $M_{\rm A47}$ residing 
in the subdiagram $S_1$ can be achieved.
[Actually, ${\sf C}_{S_1} M_{\rm A47}$ still has divergences from
the subdiagrams $S_3$ and $S_4$ which must be subtracted by other
terms of Eq.~(\ref{eq:subtraction}).]


The {\sf K}-operator, ${\sf K}_{S_4}$, 
acts on the regularized integrand $J(z)$ of $M_{\rm A47}$ directly 
and produces 
a function $J_{S_4}(z)$ that possesses the same UV singularity 
associated with the subdiagram $S_4$ as $J(z)$.
 By definition ${\sf K}$-operation also has 
the factorization property.
 For instance, the operator ${\sf K}_{S_4}$ acting 
on the regularized $M_{A47}$ 
produces the factorized result
\begin{eqnarray}
 {\sf K}_{S_4} M_{A47} &=&  \,L_2^{\rm UV}\,M_{8A}\,,
\end{eqnarray}
where $L_2^{\rm UV}$ is the UV-divergent part of 
the regularized second-order on-shell vertex renormalization constant $L_2$ 
and does not include the IR-divergent part of $L_2$
\cite{Cvitanovic:1974sv}.
 $M_{8A}$ denotes the amplitude 
of the magnetic moment from the
eighth-order diagram $A$ of Fig.~\ref{fig:8thCILbyL}.
 The regularization mass must be sent to infinity after 
${\sf K}_{S_4} M_{A47}$ is combined with $M_{A47}$.
 The difference of $L_2^{\rm UV}$ and $L_2$ is
accounted for at the stage of the residual renormalization.

 The subtraction term
${\sf C}_{S_3} M_{\rm A47}$ can be written (somewhat) symbolically as
\begin{eqnarray}
 {\sf C}_{S_3} M_{\rm A47} &=& \Pi (0,0,0,0) M_4,
\end{eqnarray}
 Here $\Pi$ is a short-hand form of the sixth-order {\it l-l} subdiagram
defined by
\begin{eqnarray}
\Pi_{\kappa \lambda \mu \nu} (k_a, k_b, k_c, k_d)|_{k_a=k_b=k_c=k_d=0},
\end{eqnarray}
where $k_a$, etc., are the momenta carried by the photon line $a$, etc.,
and $M_4$ is obtained from $M_{A47}$ by shrinking the {\it l-l} loop
of $S_3$ to a point.
 The UV divergence of $M_{A47}$ arising from the subdiagram $S_3$
is cancelled by the term ${\sf C}_{S_3} M_{\rm A47}$ 
of Eq.~(\ref{eq:subtraction}), which results in full renormalization
of the $S_3$ divergence.
 Actually, $M_{A47}$ contains another UV divergence arising from $S_4$
which we subtract by the operator $K_{S_4}$.
The complete removal of UV divergences arising from $S_3$ and $S_4$
is achieved by the combination
\begin{eqnarray}
  M_{A47} -{\sf C}_{S_3} M_{A47} 
   -{\sf K}_{S_4} M_{A47} 
   +{\sf K}_{S_4} {\sf C}_{S_3} M_{A47}.
\end{eqnarray}

 When the contributions of all diagrams of Set II(e)
listed in Eq.~(\ref{eq:indpndntDiag_setIIe})
are put together,
$\Pi (0,0,0,0)$ from all diagrams cancel out and we obtain a simple result
\begin{eqnarray}
 A_l({\rm II(e)},\,l^\prime) &=& 
 \sum_{Gij \in {\rm\,Eq.(\ref{eq:indpndntDiag_setIIe})}} 
 \Delta M_{Gij}
 - 4 
 \left\{ 
  \left( L_2 - L_2^{\rm UV} \right)
  +
  \left( B_2 - B_2^{\rm UV} \right)
 \right\}
 A_l({\rm LL8},\,l^\prime)\nonumber\\
 &=&
 \sum_{Gij \in {\rm\,Eq.(\ref{eq:indpndntDiag_setIIe})}} 
 \Delta M_{Gij}
 - 4 \Delta B_2 \times A_l({\rm LL8},\,l^\prime)\,,  \nonumber \\
 & & \Delta B_2 \equiv \left\{ 
  \left( L_2 - L_2^{\rm UV} \right)
  +
  \left( B_2 - B_2^{\rm UV} \right)
 \right\} = \frac{3}{4}.
  \label{eq:finite_renormalization}
\end{eqnarray}
Now, at last, we can send the regulator mass to infinity.
$B_2$ is the full second-order wave function renormalization constant
and $B_2^{\rm UV}$ is the UV-divergent part of $B_2$ defined by
the ${\sf K}$ operation.
 The IR divergence of $(L_2 - L_2^{\rm UV})$ 
cancels that of $(B_2 - B_2^{\rm UV})$ 
leaving a finite term as expected.
 (Note that $B + L = 0$ while $B_2^{\rm UV} + L_2^{\rm UV}$ is finite 
but not zero. 
See Ref.~\cite{Cvitanovic:1974um} for the exact definitions of 
$B_2^{\rm UV}$ and $L_2^{\rm UV}$.)
 Eq.~(\ref{eq:finite_renormalization}) shows that 
the residual renormalization term is proportional to
the eighth-order contribution $A_l({\rm LL8},\,l^\prime)$
to the anomalous magnetic moment of the lepton $l$
from the diagrams of Fig.~\ref{fig:8thCILbyL}, 
in which loops are given by lepton $l^\prime$.
 The numerical study in Ref.~\cite{Kinoshita:2002ns} 
has provided an accurate value for the mass-independent contribution
$A_e({\rm LL8},\,e)$ 
\begin{eqnarray}
 A_e({\rm LL8},\,e) = -0.990\ 72\ (10)\,. \label{eq:res_ee}
\end{eqnarray}
 In addition the paper~\cite{Kinoshita:2002ns} reports 
the electron-loop contribution to $a_\mu$
\begin{eqnarray}
 A_\mu({\rm LL8},\,e) &=& -4.432\ 43\ (58) \,.\label{eq:res_muonE} 
\end{eqnarray}
 We have also evaluated the muon loop contribution 
$A_e({\rm LL8},\,\mu)$ to the electron $g\!-\!2$, 
and the tau-lepton loop contribution $A_\mu({\rm LL8},\,\tau)$
to the muon $g\!-\!2$ needed for this work
\begin{eqnarray}
 A_e({\rm LL8},\,\mu) &=& -0.000~177~8~(12) ,
  \label{eq:res_eMu} \\
 A_\mu({\rm LL8},\,\tau) &=& -0.015~868~(37).
  \label{eq:res_muTau} 
\end{eqnarray}
 The remaining task is to evaluate 
every $\Delta M_{Gij}$ 
in various combination of the external and internal leptons.

\section{Numerical results of $\Delta M_{Gij}$}
\label{sec:int_renorm}

{\sc fortran} codes for $\Delta M_{Gij}$ 
are rather complicated and not easy to obtain.
In order to facilitate this problem
 we adapted the automating code 
{\sc gencode}{\it N} specifically for the Set II(e) which generates
the integrands of $\Delta M_{Gij}$ 
as {\sc fortran}-formatted source programs.
(See Refs.~\cite{Aoyama:2005kf,Aoyama:2007bs} for the details of automation.)
 Two independent sets of automating codes 
together with another set of manually-produced codes were constructed 
to confirm their validity.

 The integral for the diagram $Gii$, 
i.e, the one containing the second-order self-energy subdiagram, 
was found to exhibit worse convergence than the others.
In order to alleviate this problem, we modify the integrand
in the following way.
 For instance, in the diagram $A44$ in Fig.~\ref{fig:setII(e)}, 
which contains a second-order self-energy subdiagram,
 the integrand of $\Delta M_{Gij}$ 
depends on the Feynman parameters $z_4,\,z_8$ 
only through the combination $z_{48} \equiv (z_4 + z_8)$.
 Thus, the number of independent variables is reduced from 12 to 11.
 This seems to improve somewhat the convergence of iteration procedure.

The integration of $\Delta M_{Gij}$ 
is carried out with the help of the adaptive-iterative 
Monte-Carlo integration routine VEGAS 
\cite{vegas:Lepage} 
on the massively parallel computer,
RIKEN Integrated Cluster of Clusters (RICC).
The number of sampling points for each iteration is
$10^8$ for all diagrams 
with the second-order self-energy subdiagrams 
and $2 \times 10^8$ for all others.

\begin{table}[htb]
\begin{ruledtabular}
\caption{
Numerical results for mass-independent $\Delta M_{Gij}$ 
from diagrams $Gij$ in Set II(e). 
Full symmetry factors are included for the individual values. 
The number of sampling points for each iteration is
$10^8$ for all diagrams 
with the second-order self-energy subdiagrams 
and $2 \times 10^8$ for all others.
The first, second, third and fourth columns show the name
of the diagram, 
the value of integrand and its uncertainty, the $\chi^2$ value of 
VEGAS integration,
and the number of iterations of VEGAS integration.
If $\chi^2$ is very close to 1, then the numerical
integration by VEGAS is reliable.
\label{tab:IIe_deltaM_ee}
}
\begin{tabular}{@{\hskip2em}l@{\hskip-2em}ddr@{\hskip2em}} 
\multicolumn{1}{r}{ {\hskip2em}$Gij$} & 
\multicolumn{1}{r}{ $\Delta M_{Gij}$\ (uncertainty)  } &
\multicolumn{1}{c}{ $\chi^2$  } & 
\multicolumn{1}{c}{ {\hskip-2em}iterations} \\ 
\hline 
$A44$ &   5.397\ 41\ (33)  & 1.04 & 2420\\
$A55$ &   2.796\ 88\ (23)  & 1.09 & 1250\\
$A77$ &   2.422\ 84\ (20)  & 1.08 & 1210\\
$A47$ &   0.100\ 26\ (15)  & 1.02 & 1120\\
$A45$ &  -1.559\ 22\ (16)  & 0.98 & 1150\\
$A46$ &   1.180\ 76\ (12)  & 1.10 & 1140\\
$A57$ &   1.653\ 245\ (93) & 1.09 &  800\\
\hline
$B44$ &  -4.440 \ 95\ (34) & 1.06 & 2660\\
$B55$ &  -4.741\ 06\ (33)  & 0.99 & 2500\\
$B47$ &   1.725\ 96\ (17)  & 1.05 &  990\\
$B45$ &   2.521\ 96\ (17)  & 1.07 & 1070\\
$B46$ &  -0.349\ 57\ (11)  & 1.04 & 1040\\
$B57$ &  -2.254\ 206\ (97) & 1.10 &  790\\
\hline
$C44$ &  -5.054\ 64\ (34)  & 1.03 & 2570\\
$C55$ &  -2.398\ 68\ (24)  & 1.09 & 1670\\
$C77$ &  -2.431\ 20\ (22)  & 1.06 & 1460\\
$C47$ &   1.574\ 62\ (16)  & 1.06 &  990\\
$C45$ &   1.821\ 84\ (17)  & 1.00 & 1200\\
$C46$ &  -1.881\ 49\ (12)  & 1.03 & 1280\\
$C57$ &  -0.401\ 777\ (91) & 1.03 &  710\\
\hline
sum &  -4.317\ 02\ (94) & & \\
\end{tabular} 
\end{ruledtabular}
\end{table} 
%

\subsection{electron $g\!-\!2$}
\label{subsec:electron_g-2}

The results of numerical integration of $\Delta M_{Gij}$ 
for the mass-independent 
contribution to the lepton $g\!-\!2$ 
are presented in Table \ref{tab:IIe_deltaM_ee}.
Following Eq.~(\ref{eq:finite_renormalization}) 
the last line of Table \ref{tab:IIe_deltaM_ee} together 
with the value (\ref{eq:res_ee}) for $A_e({\rm LL8},\,e)$
yields the mass-independent contribution to $g\!-\!2$ 
from Set II(e) diagrams
\begin{eqnarray}
 A_l({\rm II(e)},\,l) &=& -1.344\ 86\ (99) ,
  \label{eq:ans_ee}
\end{eqnarray}
where $l = e,\mu$, or $\tau$.
Recall that the 
actual contribution to $g\!-\!2$, $a_e({\rm II(e)},\,e)$, 
is $A_e({\rm II(e)},\,e)$ times 
the factor $\left(\frac{\alpha}{\pi}\right)^5$.

\begin{table}[htb] 
\begin{ruledtabular}
\caption{Numerical results for $\Delta M_{Gij}$ of electron $g\!-\!2$ 
from diagrams $Gij$ in set II(e) 
in each of which muon induces light-by-light scattering.
 Full symmetry factors are included in the individual values. 
The number of sampling points for each iteration is
$10^8$ for all diagrams 
with the second-order self-energy subdiagrams 
and $2 \times 10^8$ for all others.
\label{tab:IIe_deltaM_eMu} 
}
\begin{tabular}{@{\hskip2em}l@{\hskip-2em}ddr@{\hskip2em}} 
\multicolumn{1}{r}{ {\hskip2em}$Gij$} & 
\multicolumn{1}{r}{ $\Delta M_{Gij}$ (uncertainty) $\times 10^3$  } &
\multicolumn{1}{c}{ $\chi^2$  } & 
\multicolumn{1}{c}{ {\hskip-2em}iterations} \\ 
\hline 
$A44$  & 11.424\ 89\ (49) & 1.19 & 1360\\
$A55$  &  5.822\ 60\ (36) & 1.29 &  640\\
$A77$  &  6.167\ 18\ (33) & 1.23 &  640\\
$A47$  &  3.674\ 56\ (28) & 1.06 &  320\\
$A45$  &  0.715\ 82\ (275)& 1.14 &  320\\
$A46$  &  3.296\ 85\ (25) & 1.18 &  400\\
$A57$  &  3.686\ 56\ (20) & 1.55 &  240\\
\hline
$B44$  & -6.068\ 57\ (45) & 1.02 & 1280\\
$B55$  & -7.947\ 942\ (44)& 0.98 & 1280\\
$B47$  & -1.252\ 29\ (29) & 1.16 &  240\\
$B45$  &  0.775\ 54\ (25) & 1.04 &  320\\
$B46$  & -0.389\ 63\ (26) & 0.91 &  240\\
$B57$  & -3.590\ 13\ (20) & 1.54 &  240\\
\hline
$C44$  &  -7.361\ 42\ (46) & 1.14 & 1120\\
$C55$  &  -3.125\ 44\ (31) & 1.17 &  720\\
$C77$  &  -2.927\ 29\ (34) & 1.33 &  480\\
$C47$  &   0.213\ 13\ (28) & 0.95 &  240\\
$C45$  &  -0.306\ 03\ (30) & 0.91 &  240\\
$C46$  &  -3.423\ 30\ (26) & 1.29 &  400\\
$C57$  &  -0.383\ 70\ (17) & 1.25 &  240\\
\hline
sum &        -0.998\ 6\ (14) & &\\
\end{tabular} 
\end{ruledtabular}
\end{table} 

 The electron $g\!-\!2$ also receives the Set II(e) contribution
induced by the virtual muon loop.
 To see its numerical significance, the computation of 
$\Delta M_{Gij}$ for the muon loop is explicitly performed.
 The results are shown in Table \ref{tab:IIe_deltaM_eMu}.
Putting together the last line of this table
and the value (\ref{eq:res_eMu}) of $A_e({\rm LL8},\,\mu)$
we obtain the muon loop contribution to the electron $g\!-\!2$ 
from Set II(e) diagrams 
\begin{eqnarray}
 A_e({\rm II(e)},\,\mu) &=& -0.000~465~(4).
  \label{eq:ans_eMu}
\end{eqnarray}

 The size of this contribution is less than the numerical uncertainty of 
the electron-loop contribution $A_e({\rm II(e)},\,e)$
given in Eq.~(\ref{eq:ans_ee}).
 Since the tau-lepton loop contribution is expected to be 
about two-orders of magnitudes smaller than the muon loop contribution
and hence negligible,
we present the sum of Eqs.~(\ref{eq:ans_ee}) and (\ref{eq:ans_eMu}) 
as our current best value for the contribution to the electron $g\!-\!2$ 
from Set II(e) diagrams 
\begin{eqnarray}
 a_e({\rm II(e)})
 &=& - 1.345\ (1)  \left(\frac{\alpha}{\pi}\right)^5\,.
 \label{eq:electron_g-2_IIe}
\end{eqnarray}

\subsection{muon $g\!-\!2$}
\label{subsec:muon_g-2}

\begin{table}[htb] 
\begin{ruledtabular}
\caption{Numerical results for $\Delta M_{Gij}$ of muon $g\!-\!2$ 
from diagrams $Gij$ in set II(e)
in each of which electron induces light-by-light scattering.
 Full symmetry factors included in the individual values. 
The number of sampling points for each iteration is
$10^8$ for all diagrams 
with the second-order self-energy subdiagrams 
and $2 \times 10^8$ for all others.
\label{tab:IIe_deltaM_mue} 
}
\begin{tabular}{@{\hskip2em}l@{\hskip-2em}ddr@{\hskip2em}} 
\multicolumn{1}{r}{ {\hskip2em}$Gij$} & 
\multicolumn{1}{r}{ $\Delta M_{Gij}$\ (uncertainty)  } &
\multicolumn{1}{c}{ $\chi^2$  } & 
\multicolumn{1}{c}{ {\hskip-2em}iterations} \\ 
\hline 
$A44$ &  21.914\ 7\ (35) & 1.02 & 2180\\
$A55$ &  10.747\ 4\ (21) & 0.99 & 1540\\
$A77$ &  10.438\ 3\ (20) & 0.99 & 1300\\
$A47$ &   6.166\ 8\ (27) & 1.01 & 1370\\
$A45$ &   7.843\ 7\ (28) & 1.04 & 1370\\
$A46$ & -13.679\ 5\ (14) & 0.98 & 1050\\
$A57$ & -14.181\ 8\ (11) & 1.09 &  810\\
\hline
$B44$ & -25.919\ 5\ (40) & 1.02 & 3050\\
$B55$ & -25.634\ 7\ (37) & 1.03 & 3050\\
$B47$ &  39.794\ 5\ (28) & 1.08 & 1760\\
$B45$ &  41.011\ 0\ (28) & 1.11 & 1810\\
$B46$ & -16.936\ 4\ (15) & 0.99 & 1130\\
$B57$ & -22.029\ 9\ (11) & 1.23 &  840\\
\hline
$C44$ & -41.123\ 9\ (42) & 0.99 & 3010\\
$C55$ & -20.500\ 5\ (29) & 1.02 & 1770\\
$C77$ & -20.929\ 8\ (29) & 1.05 & 1610\\
$C47$ &  46.365\ 7\ (30) & 1.05 & 1810\\
$C45$ &  47.994\ 0\ (30) & 1.04 & 1890\\
$C46$ & -22.236\ 3\ (15) & 1.06 & 1130\\
$C57$ & -19.136\ 5\ (12) & 1.08 &  810\\
\hline
sum &        -10.032\ (12) & & \\
\end{tabular}
\end{ruledtabular}
\end{table} 

\begin{table}[htb] 
\begin{ruledtabular}
\caption{Numerical results for $\Delta M_{Gij}$ of muon $g\!-\!2$ 
from diagrams $Gij$ in set II(e) 
in each of which tau-lepton induces light-by-light scattering.
 Full symmetry factors are included in the individual values. 
The number of sampling points for each iteration is
$10^8$ for all diagrams 
with the second-order self-energy subdiagrams 
and $2 \times 10^8$ for all others.
\label{tab:IIe_deltaM_muTau} 
}
\begin{tabular}{@{\hskip2em}l@{\hskip-2em}ddr@{\hskip2em}} 
\multicolumn{1}{r}{ {\hskip2em}$Gij$} & 
\multicolumn{1}{r}{ $\Delta M_{Gij}$\ (uncertainty)  } &
\multicolumn{1}{c}{ $\chi^2$  } & 
\multicolumn{1}{c}{ {\hskip-2em}iterations} \\ 
\hline 
$A44$ &  0.422\ 177\ (18) & 1.09 & 1600\\
$A55$ &  0.215\ 057\ (18) & 1.19 &  480\\
$A77$ &  0.219\ 086\ (19) & 1.18 &  320\\
$A47$ &  0.112\ 235\ (14) & 0.96 &  240\\
$A45$ & -0.008\ 963\ (17) & 1.07 &  160\\
$A46$ &  0.119\ 699\ (16) & 1.37 &  160\\
$A57$ &  0.138\ 883\ (10) & 1.63 &  160\\
\hline
$B44$ & -0.239\ 377\ (18) & 1.02 & 1400\\
$B55$ & -0.298\ 375\ (18) & 0.99 & 1280\\
$B47$ & -0.017\ 671\ (16) & 1.11 &  160\\
$B45$ &  0.053\ 945\ (16) & 1.16 &  160\\
$B46$ & -0.014\ 195\ (13) & 0.96 &  160\\
$B57$ & -0.140\ 250\ 5\ (80) & 1.43 &  240\\
\hline
$C44$ & -0.288\ 614\ (18) & 1.04 & 1280\\
$C55$ & -0.125\ 456\ (18) & 1.14 &  400\\
$C77$ & -0.120\ 731\ (18) & 1.18 &  320\\
$C47$ &  0.019\ 880\ (15) & 1.13 &  160\\
$C45$ &  0.011\ 773\ (16) & 0.89 &  160\\
$C46$ & -0.129\ 852\ (14) & 1.09 &  240\\
$C57$ & -0.014\ 913\ 9\ (86) & 1.17 &  160\\
\hline
sum & -0.085\ 66\ (7)  & & \\
\end{tabular} 
\end{ruledtabular}
\end{table} 

 The main contribution of Set II(e) to the muon $g\!-\!2$ arises from
the diagrams each of which is induced by an electron loop.
 We present the numerical result 
of $\Delta M_{Gij}$ for the electron-loop contribution 
in Table \ref{tab:IIe_deltaM_mue}.
This table shows that the sum of $\Delta M_{Gij}$ is an order of magnitude
larger than that of the mass-independent $\Delta M_{Gij}$ 
in Table \ref{tab:IIe_deltaM_ee}.
 However,  as is seen from
Eq.~(\ref{eq:res_muonE}) for $A_\mu({\rm LL8},\,e)$, 
partial cancellation takes place between 
the first term 
and the second term [$-4 \Delta B_2 \times A_\mu({\rm LL8},\,e)$].
 As a consequence, we have 
\begin{eqnarray}
 A_\mu({\rm II(e)},\,e) &=& 3.265\ (12) \,.
  \label{eq:ans_muonE}
\end{eqnarray} 

 Thus the electron-loop contribution $A_\mu({\rm II(e)},\,e)$
is not much larger than the muon loop contribution 
$A_\mu({\rm II(e)},\,\mu)$ of Eq.~(\ref{eq:ans_ee}).
 Since the sign of $A_\mu({\rm II(e)},\,e)$ is opposite to
that of $A_\mu({\rm II(e)},\,\mu)$, we are curious about
the role that the tau-lepton contribution 
$A_\mu({\rm II(e)},\,\tau)$ might play.
 Table \ref{tab:IIe_deltaM_muTau} shows the result of $\Delta M_{Gij}$ 
for the Set II(e) contribution to the muon $g\!-\!2$ 
induced by the tau-lepton loop.
 Equation~(\ref{eq:finite_renormalization}),
together with the value (\ref{eq:res_muTau}) of $A_\mu({\rm
LL8},\,\tau)$, 
gives 
\begin{eqnarray}
 A_\mu({\rm II(e)},\,\tau) &=& -0.038\ 06\ (13) \,,
 \label{eq:ans_muTau}
\end{eqnarray}
which is two orders of magnitude smaller than 
$A_\mu({\rm II(e)},\,e)$ or $A_\mu({\rm II(e)},\,\mu)$.
 Summing up Eqs.~(\ref{eq:ans_muonE}), (\ref{eq:ans_ee}) 
and (\ref{eq:ans_muTau}), the Set II(e) contribution to the muon $g\!-\!2$
is found to be
\begin{eqnarray}
 a_\mu({\rm II(e)}) &=& 
 1.882\ (13)  \left(\frac{\alpha}{\pi}\right)^5\,.
 \label{eq:muon_g-2_IIe}
\end{eqnarray}

\section{discussion and summary}
\label{sec:conclusion}

 In this paper, we computed the contribution to the lepton $g\!-\!2$ 
from the tenth-order QED diagrams of Set II(e) that contain 
the sixth-order light-by-light-scattering subdiagram
internally. 
The use of Ward-Takahashi identity,
as well as the symmetries of QED, 
reduces the computation of 180 Feynman diagrams 
to that of 20 integrals $\Delta M_{Gij}$.
 The intermediate renormalization to define $\Delta M_{Gij}$ 
is chosen so that the UV divergence associated 
with the second-order self-energy or vertex subdiagram 
is subtracted away by ${\sf K}$-operation. 
Meanwhile the UV divergence arising from the {\it l-l} loop
is subtracted by full renormalization.
 This leads to simplification of the final result
as is seen in Eq.~(\ref{eq:finite_renormalization}).

 The Set II(e) contribution to the electron $g\!-\!2$ 
is obtained by evaluating the electron and muon virtual effects.
The result is given in Eq.~(\ref{eq:electron_g-2_IIe}).
 The size is of the typical order of magnitude 
for the tenth-order.
The numerical computation was carried out as accurately as possible 
with the available computer resources.

 The contribution to the muon $g\!-\!2$ is obtained
by evaluating the virtual effects of all leptons. 
The result is given in Eq.~(\ref{eq:muon_g-2_IIe}).
 The contribution of the electron loop to the muon $g\!-\!2$ 
is not much larger than the muon loop contribution.

 We found that $A_\mu({\rm II(e)},\,e)$ in Eq.~(\ref{eq:ans_muonE}) 
involves partial cancellation between the sum of 
$\Delta M_{Gij}$ over all $Gij$ in Set II(e) 
and the residual renormalization term in Eq.~(\ref{eq:finite_renormalization}).
 In spite of these problems we were able to obtain the result
for $a_\mu(II(e))$ with the uncertainty less than $1\%$
using the high performance computer system, RICC. 

\begin{acknowledgments}
 This work is supported in part by 
JSPS Grant-in-Aid of Scientific Research (C) Grants No.~19540322
and No.~20540261, 
and Grant-in-Aid of Ministry of Education Grant No.~20105002.
The work of T. K. was supported by the U. S. National Science Foundation
under Grant No.~NSF-PHY-0757868.
T. K. thanks RIKEN for the hospitality extended to him where
a part of this work was carried out.
 Numerical computation was mostly conducted on 
the RIKEN Integrated Cluster of Clusters (RICC).
 A part of preliminary computation was also conducted 
on the computers of the theoretical particle physics group (E-ken), 
Nagoya University.
\end{acknowledgments}


\end{document}